# Image enhancement algorithm for absorption imaging

**Pengcheng Zheng (郑鹏程)**[1,2,†], **Songqian Zhang (张松骞)**[1,†], **Zhu Ma (马翥)**[1,2], **Haipo Niu (牛海坡)**[1], **Jiatao Wu (吴嘉涛)**[1,2], **Zerui Huang (黄泽锐)**[1], **Chengyin Han (韩成银)**[2], **Bo Lu (鹿博)**[2], **Peiliang Liu (刘培亮)**[1**] and **Chaohong Lee(李朝红)**[2*]

[1]Laboratory of Quantum Engineering and Quantum Metrology, School of Physics and Astronomy, Sun Yat-Sen University (Zhuhai Campus), Zhuhai 519082, China

[2]Institute of Quantum Precision Measurement, State Key Laboratory of Radio Frequency Heterogeneous Integration, College of Physics and Optoelectronic Engineering, Shenzhen University, Shenzhen 518060, China

[†]These authors contributed equally to this work and should be considered co-first authors

*Corresponding author: chleecn@szu.edu.cn; ** Corresponding author: liupliang@mail.sysu.edu.cn

The noise in absorption imaging of cold atoms significantly impacts measurement accuracy across a range of applications with ultracold atoms. It is crucial to adopt an approach that offers effective denoising capabilities without compromising the unique structure of the atoms. Here we introduce a novel image enhancement algorithm for cold atomic absorption imaging. The algorithm successfully suppresses background noise, enhancing image contrast significantly. Experimental results showcase that this approach can enhance the accuracy of cold atom particle number measurements by approximately tenfold, all while preserving essential information. Moreover, the method exhibits exceptional performance and robustness when confronted with fringe noise and multi-component imaging scenarios, offering high stability. Importantly, the optimization process is entirely automated, eliminating the need for manual parameter selection. The method is both compatible and practical, making it applicable across various absorption imaging fields.

**Keywords**: cold atoms; absorption images; denoising algorithm; self-adaptive gauss filter; minimal description length

1. Introduction.

Experiments of ultracold atoms serve as a valuable platform for studying precision measurement [1-3] and many-body quantum phenomena [4,5]. In these investigations, the atomic distribution is typically evaluated using light absorption images [6-8] taken at a predetermined atomic interaction time. A precise determination of the atomic distribution can provide crucial information such as temperature, number, and density of atoms, which is essential for applications in quantum metrology [9], the measurement of physical parameters [10,11], and the study of phase transitions or dimensional crossover [12-13]. Accurately measuring the atomic cloud distribution is particularly crucial in fields like interferometry [14-15] and vortex couplers [16-19], where it serves as a critical component.

The traditional absorption imaging procedure involves taking three photos in a row as shown in figure1: the image L from the atomic cloud, the image G to record the light distribution, and the image N to capture the background signal. The optical density A(x,y) in the x-y plane is represented by these three images

$$A(x,y) = -\log \frac{L_R(x,y) - N(x,y)}{G_R(x,y) - N(x,y)} \quad (1)$$

where the subscript $R$ on $L_R$ and $G_R$ indicates that these are the actual taken photos, not the datas used in the Numerical calculation. Therefore, given the cross-section $\sigma$, $A/\sigma$ gives the two-dimensional atomic distribution. In theory, the atomic cloud alone is the sole factor causing the difference between L' and G'. However, in practice, noise from the camera during the measurement process accounts for the majority of the error, which is hard to remove by adjusting the optical route.

In this article, we explore the noise and grayscale characteristics of absorption imaging of cold atoms and introduce a cold atomic absorption imaging enhancement algorithm. The algorithm consists of two components: adaptive Gaussian filtering and nonlinear gray stretching. The former focuses on denoising the atomic cloud signal using the minimum description length concept. Subsequently, the nonlinear gray-dot calculation effectively removes background noise and enhances image contrast. Experimental results show that this approach can enhance particle number measurement accuracy by approximately tenfold without compromising the unique structure of the atoms. Moreover, even when dealing with fringe noise and multi-component imaging, the method exhibits exceptional efficacy and high stability, with the optimization process being fully automated without the need for manual parameter selection. These advantages position the algorithm to better support applications in quantum control, precision measurement, and many-body quantum physics.

2. Theory.

CMOS or CCD solid-state photosensors are widely used in contemporary measurement instruments to transform light into digital signals. Such a conversion is not optimal due to the flaws in photosensors and results in noise in the measured signal.

Generally speaking, the following linear model may be used to describe the production of a digital sensor raw picture D [20],

$$D = KI + N \quad (2)$$

where $I$ is the number of photoelectrons proportional to scene irradiation, $K$ is the overall system gain composed of analog and digital gains, and $N$ denotes the sum of all noise sources physically caused by light or camera.

Figure 1 depicts the current dominant imaging photosensor. To simulate noise, we investigate the electronic imaging pipeline of the photoelectric conversion from photons to electrons, electrons to voltage, and voltage to digital values.

During exposure, incident light in the form of photons strikes the photosensor pixel area, releasing photon-generated electrons (photoelectrons) proportional to the light intensity. The quantity of electrons captured will always be unpredictable because of the quantum nature of light. This quantity of electrons is subject to a Poisson distribution,

$$(I + N_p) \sim \mathcal{P}(I) \quad (3)$$

where $\mathcal{P}$ stands for the Poisson distribution and $N_p$ is referred to as the photon shot noise, I is the number of photoelectrons proportional to scene irradiation. The light intensity determines this type of noise. Even with a flawless sensor, shot noise is a basic constraint that cannot be avoided. Other noise sources that are introduced during the photon-to-electron step include light response nonuniformity and dark current noise $N_d$, which have been extensively documented in earlier research [21-24].

At the conclusion of the exposure time, electrons are generally integrated, amplified, and read out as quantifiable charge or voltage at each location. The noise present during the electron-to-voltage step is determined by the circuit design and processing technique employed, and is thus referred to as pixel circuit noise [21]. Thermal noise, reset noise [25], source follower noise [26] and banding pattern noise [21] are all included. The physical origins of these noise components may be discovered in the literature on electronic imaging [22,23,25,26]. For example, source follower noise is caused by traps in the silicon lattice that randomly catch and release carriers, whereas banded pattern noise is caused by the CMOS circuit readout pattern and the amplifier. Here, we take into account the thermal noise $N_t$, source follower noise $N_s$, and banding pattern noise $N_b$, and combine numerous noise sources into a single term, readout noise.

$N_{read} = N_b + N_t + N_s$  (4)

The distribution of readout noise can be approximated as Gaussian distribution

$N_{read} \sim N(\mu, \sigma^2)$  (5)

where $\mu$ and $\sigma^2$ indicate the expectation and variance respectively. The analog voltage signal read out during the last stage is quantized into discrete codes by an ADC to create a picture that may be stored on a digital storage device. This procedure introduces quantization noise $N_q$, which is given by

$N_q \sim U(-1/2q, 1/2q)$  (6)

where $U(\cdot,\cdot)$ denotes the uniform distribution over the range $[-1/2q, 1/2q]$ and $q$ is the quantization step.

To summarize, our noise formation model consists of four major noise components:

$N = KN_p + N_d + N_{read} + N_q$  (7)

where $K$, $N_p$, $N_d$, $N_{read}$ and $N_q$ denotes the overall system gain, photon shot noise, dark current noise, readout noise and quantization noise, respectively.

Since the inaccuracy induced by quantization noise is generally within half a pixel value, it is ignored by standard denoising methods. We can define readout noise and dark current noise by obtaining long-exposure photographs without light and removing them by subtraction [27-28]. A Gaussian filter may often be used to denoise photon-shot noise [29]. Spatial filtering techniques, such as the Gaussian filter, are frequently employed [30-31] due to their exceptional noise reduction performance and low computational cost.

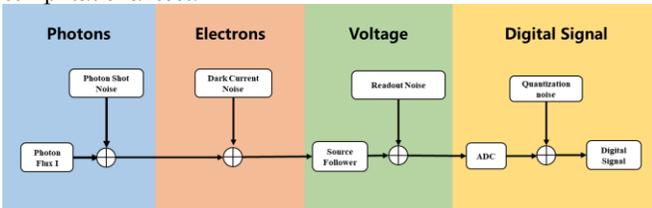

Fig. 1. Overview of electronic imaging pipeline and visualization of noise sources.

Absorption imaging includes sending a near-resonant light onto an ultracold atom cloud and capturing the "shadow" absorbed by the atoms with the camera. The measurement of this "shadow" reveals the distribution of the atomic cloud. After data processing (as in equation (1)), the "shadow" corresponding to the atomic cloud will be turned into a bright region, and the area lighted by the near resonance light will become a random noise of lesser brightness.

Consequently, brightness values—also known as gray values—have crucial characteristics for absorption imaging. On the one hand, the absorption effect will make the cool atom cloud to seem brighter; background noise, on the other hand, correlates to a reduced brightness since the area is unable to absorb the detected light's energy. As a result, the difference in brightness levels in the image allows us to differentiate between signal and noise. To be specific, areas with high brightness often represent signals, and areas with low brightness represent background noise.

3. Design of denoising algorithm.

On the basis of the above study, we propose an image enhancement method, shown in Figure 2. First of all, for the results of absorption imaging in Figure 2 (A), the SNR is mainly affected by shooting noise during camera shooting, which cannot be eliminated by optimizing the optical path. For this type of noise, we can use Gaussian filtering to get a lower noise level and higher contrast results (as shown in Figure 2 (B)). On this basis, we distinguish signal from noise by image brightness, and use a nonlinear gray transform to preserve atomic cloud and suppress background noise. The result can be a low-noise, high-contrast image of the atomic clouds (see Figure 2 (C)). Each phase is detailed below.

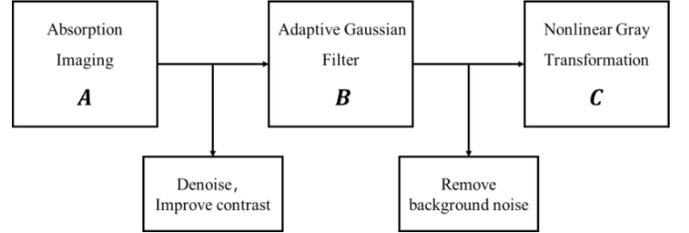

Fig. 2. Overview of electronic imaging pipeline and visualization of noise sources.

Gaussian filtering, which is the convolution of a two-dimensional image $I(x,y)$ with the two-dimensional Gaussian function, is a fundamental image processing technique used for image denoising [32].

$(G_\sigma * I)(x,y) = \iint G_\sigma(x-u, y-v)I(u,v)dudv$  (8)

$G_\sigma(x,y) = G_\sigma(x) \cdot G_\sigma(y)$  (9)

$G_\sigma(x) = \frac{1}{\sigma\sqrt{2\pi}} e^{-\frac{x^2}{2\sigma^2}}$  (10)

where $G_\sigma(x,y)$ is the Gaussian kernel and $\sigma^2$ is the variance, which determines the shape of the Gaussian kernel.

The degree to which the Gaussian filter modifies the image depends on the value of σ. When σ is large, the Gaussian filter gray makes the image smoother, which will help to denoise, but at the same time, it will destroy the original information of the image, making some features lost. Conversely, when σ is small, noise reduction is likewise ineffective, but information protection is enhanced. As a result, processing every pixel in a picture using Gaussian filtering with a single σ is difficult as doing so necessitates making a trade-off between conserving information and denoising. The adaptation of the Gaussian kernel shape to the picture structure is a natural extension of Gaussian smoothing [33]. In other words, in order to conserve details while also remove noise, the proper local variance $\sigma^2$ must be determined based on the local characteristics of the image.

The estimated local variance must be suitable for at least two basic requirements: noise reduction and feature preservation. In other words, we need to minimize residuals while maximizing the denoising. Residuals represent the difference between the original image and the processed image, and the smaller the residuals, the higher the degree of information protection for the original image. In reality, the minimal description length principle (MDL) introduced by Rissanen makes it simple to determine the best code [34-36]. The optimal local variance may be chosen using the following rule [37]:

$$\sigma_{best} = [argmin_\sigma \frac{c}{\sigma^2} + \varepsilon^2] \quad (11)$$

where $c = 4.0 \times 10^{-3} l^2$, $l$ is the edge lengths of the mesh. Residual $\varepsilon$ can be seen as [38]:

$$I_0(x,y) = I_\sigma(x,y) + \varepsilon(x,y) \quad (12)$$

where $I_0$ is the original image minus and $I_\sigma$ is the smoothed image, $I_\sigma = I_0 * G_\sigma$.

The concept of choosing the minimal description length has been effectively employed in machine vision to achieve a balance of simplicity and accuracy. According to the minimal description length, optimum coding uses the fewest number of bits required to maximize the information of both parameters. That is, the greatest smoothness with the minimal residual [39].

As previously discussed, in absorption imaging, it is feasible to use a Gaussian filter to reduce the shooting noise that cannot be eliminated directly by changing the optical path.

We modified the adaptive approach of the parameters on the basis of MDL to account for denoising and information protection. Figure 3(A) illustrates the full procedure. In order to find the input image's center of mass $(x_c, y_c)$ and remove its edge, the "log" operator is applied firstly, as illustrated in Figure 4(B). Crop the picture such that it is centered on $(x_c, y_c)$ (Figure 4(C)). The cross section of gray scale along the yellow dotted line in Figure 4(C) was then fitted with a Gaussian function (Figure 4(D)), and the sites of 15% and 1% of the center intensity were chosen as $r_s$ and $r_e$ respectively, as shown in Figure 4(E). Then in Figure 4(F), the circular areas $os$ and $oe$ is shown with the center of mass $(x_c, y_c)$ as the center of the circle and the radius of $r_s$ and $r_e$, respectively. The distribution of the atomic cloud is represented by area J within $os$, the background noise is represented by region K outside $oe$, and the remaining ring region is represented by region L. Local variance is governed by the following piecewise function,

$$\sigma = \begin{cases} [argmin_\sigma \frac{c}{\sigma^2} + \varepsilon^2] & 0 < r \leq r_s \\ a \cdot (r-b)^c + d & r_s < r \leq r_m \\ \sqrt{R^2 - (r-xc)^2} + yc & r_m < r \leq r_e \\ \sigma_e & r > r_e \end{cases} \quad (13)$$

The values of relevant parameters are as follows:

$$\begin{cases} r_m = \frac{r_s + r_e}{2} \\ c = \frac{(r_e - r_m)\sqrt{R^2 - (r_m - r_e)^2}}{(r_m - r_s)(\sigma_m - \sigma_s)} \\ a = (\sigma_m - \sigma_s) \cdot (r_m - r_s)^{-c} \\ b = r_s \\ d = \sigma_s \\ xc = r_m \\ yc = \frac{(r_e - r_m)^2}{2(\sigma_m - \sigma_e)} + \frac{(\sigma_m - \sigma_e)}{2} \end{cases} \quad (14)$$

The MDL is employed only in the distribution zone of the atomic cloud to get the optimum, as illustrated in Figure 4 (G), because the background noise beyond this region is the "useless" signal. The ideal distribution of $\sigma_c$ in region J (atomic cloud) will be obtained by MDL, and the minimal value of $\sigma_c$ is designated $\sigma_s$. Thus the noise can be reduced to a low level without causing significant random perturbations in Area K (background noise). Then the K change to be K' by applying Gaussian filtering with an appropriate $\sigma$ by MDL. And the mean $M_k$ and standard deviation $S_k$ of all the pixel values in the K' region after calculation will be used to evaluate the average intensity and volatility of background noise, respectively. The $\sigma$ is shown as $\sigma_e$ when $M_k \leq 0.1$ and $S_k \leq 0.05$. And the value of $\sigma$ in the region L is described by a nonlinear piecewise function for good results. After the above calculation, we can get the best value of σ for each pixel in the image. Finally, figure 4(I) can be obtained by Gaussian filtering Figure 4(C) with the σ value solved above (as shown in Figure 4(H)). The improved atomic cloud distribution has less noise while maintaining the distinctive structure. Furthermore, as compared to prior optimization, the background noise is greatly reduced, and a stronger contrast is gotten.

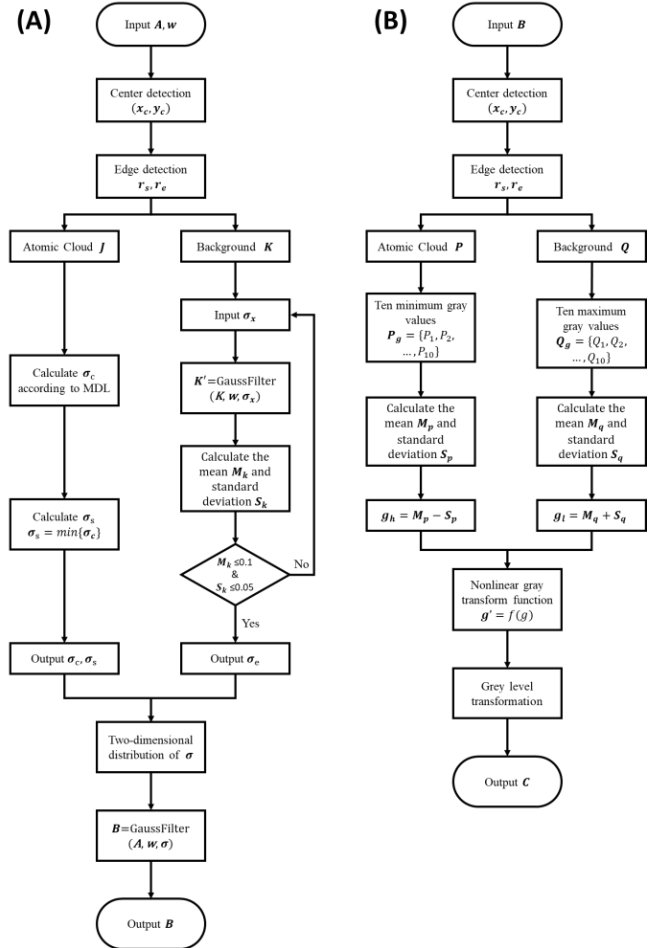

Fig. 3. Schematic diagram of Adaptive Gaussian Filter (A) and Nonlinear Gray Transform

The background noise in the absorption imaging is efficiently reduced after adaptive Gaussian filtering, so the atomic cloud has greater intensity than the background noise. To eliminate background noise further and increase image contrast, the following nonlinear piecewise function is applied to conduct point operations on the gray value of the aforementioned image.

$$g' = \begin{cases} mg^\gamma, & 0 \leq g < g_l \\ \sqrt{r^2 - (g-a)^2} + b, & g_l \leq g < g_h \\ g, & g \geq g_h \end{cases} \quad (15)$$

The values of relevant parameters are as follows:

$$\begin{cases} a = \frac{g_l^2 + 2g_h^2 - 4g_2g_1 + y_1^2}{2(x_l - y_1)} \\ b = 2g_h - a \\ r^2 = 2(g_h - a)^2 \\ \gamma = -\frac{g_l(x_1 - a)}{g_1(y_1 - b)} \\ m = \frac{y_1}{g_l^{\gamma}} \end{cases} \quad (16)$$

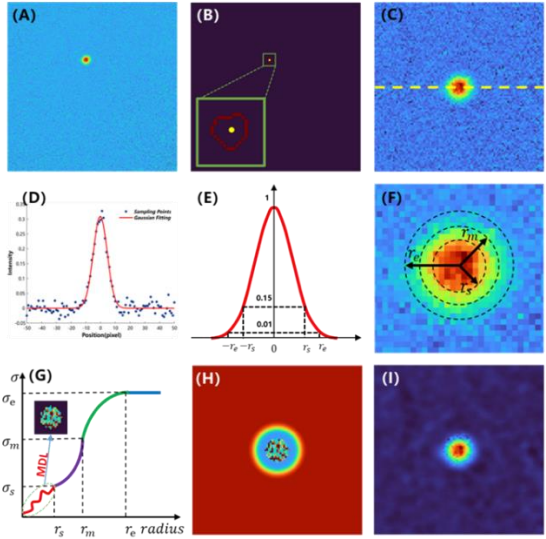

Fig. 4. Schematic diagram of Adaptive Gaussian filtering. (A) Absorption imaging with a width of 301 pixels; (B) Edge (red) and center (yellow) of the atomic cloud; (C) The image is cropped from the center of the atomic cloud and its width is 101 pixels; (D) Gaussian fitting (solid red line) of the intensity along the yellow dotted line in (C); (E) Calculation of $r_s$ and $r_e$; (F) Dividing (C) into three regions based on $r_s$, $r_e$ and $r_m$; (G) Diagram of calculation of $\sigma$ in different regions; (H) The two-dimensional distribution of σ corresponding to (C); (I) The result of adaptive Gaussian filtering of the image (C).

Holding the grayscale value constant in the image when $g$ is bigger than $g_h$, so it will keep the information of the atomic cloud, as illustrated in Figure 5. In addition, applying a gamma transform to suppress low-intensity background noise when $g$ is smaller than $g_l$. The circle function is utilized to produce a smooth transition at the fraction point, and the key point is to determine the values of $g_l$ and $g_h$. The specific process is shown in Figure 3(B).

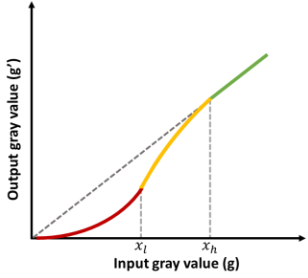

Fig. 5. Diagram of nonlinear gray transform function (equation (15)).

The atomic cloud's center position $(x_c, y_c)$ need to be figured out firstly, then the whole image is divided into three regions: the atomic cloud region P, background noise region Q, and transition region R. The procedure is the same as that used in adaptive Gaussian filtering. The ten minimum gray values $P_g$ in the region P are utilized to compute the mean $M_p$ and standard deviation $S_p$ of $P_g$, while ten maximum gray values $Q_g$ in the region Q are used to the mean $M_q$ and standard deviation $S_q$. $g_h$ and $g_l$ are the brightness values that distinguish the signal, noise, and the intermediate region.

4. Experiment and Analysis.

To evaluate the algorithm's efficacy, we crafted an atomic Bose-Einstein condensate (BEC) comprising $^{87}$Rb atoms within an asymmetric crossed optical dipole trap (ACODT). This ACODT is constructed from two focused laser beams intersecting at a 60° angle, each with variable radii to optimize trap frequency and volume. The large optical dipole trap (L-ODT) boasts an 82 µm radius, while the small optical dipole trap (S-ODT) features a 48 µm radius. Initially, the L-ODT efficiently captures a substantial quantity of atoms. During evaporation, the S-ODT provides a higher trap frequency. Our procedure commences with a standard magneto-optical trap (MOT), followed by the implementation of a temporal dark MOT to enhance atom density, before directly loading atoms into the ACODT. To maximize atom retention during the $i$-th evaporation step, we optimize both the power ratio of L-ODT to S-ODT and the evaporation time. Subsequent to evaporative cooling of laser-cooled atoms within the ACODT, we successfully generate an $^{87}$Rb BEC [40].

Absorption imaging allows us to capture the distribution of BEC, as seen in Figure 6(A), where (D) and (E) represent the cross section of the intensity distribution of (A) and the background noise gray level histogram, respectively. By contrasting (D) and (E), it is clear that adaptive Gaussian filtering may reduce the noise disruption while maintaining atomic cloud characteristics and suppressing background noise. (F) shows that the nonlinear gray transform may eliminate background noise greatly without altering the atomic cloud's spatial distribution. At the same time, we demonstrate the change throughout algorithm execution using a gray histogram. The mean and standard deviation of background noise in (G) are 0.126 and 0.032, and 0.021 and 0.005 in (H), respectively. It shows that adaptive filtering suppresses background noise by six times. And there is hardly any background noise in (H). The results above indicate that our method can eventually reduce the influence of noise on atomic clouds and nearly totally eliminate background noise.

In order to evaluate the performance of this algorithm more accurately, we further measure the full width at half maximum (FWHM) and particle number of atomic clouds at different tof times, and compare the results with or without our algorithm.

Figure 7 depicts the FWHM of an atomic cloud at various delay durations. When compared to absorption imaging, the image enhancement approach used here has no influence on atomic cloud size measurement, indicating that our algorithm does not change the spatial distribution of atomic clouds and has excellent fidelity.

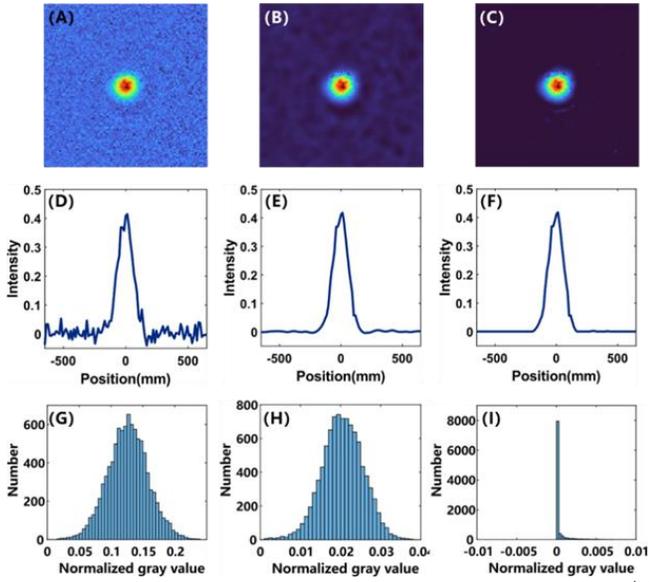

Fig. 6. Absorption imaging and image enhancement of BEC. (A): absorption imaging of BEC; (B): (A) the result of adaptive Gaussian filtering; (C): (B) the nonlinear gray transform results; (D) ~(F) : cross section of intensity distribution corresponding to (A) ~(C); (G) ~(I) : histogram of gray distribution of background noise corresponding to (A) ~(C)

The kinetic temperature of atomic cloud may be estimated using its size. Its experssion is [41],

$T = \frac{m\sigma_T^2}{2k_B t_{TOF}^2}$     (17)

where V is the atomic mass, $t_{TOF}$ is the TOF time, $k_B$ is the Boltzmann constant and $\sigma_T$ is the standard deviation from the Gaussian fitting. The relationship between FWHM and standard deviation $\sigma$ of Gaussian function is as follows [42]:

$FWHM = 2\sqrt{2 \ln 2}\, \sigma$     (18)

So the kinetic temperature of the atomic cloud can be estimated by the FWHM,

$T = \frac{mFWHM^2}{16 \ln 2 k_B t_{TOF}^2}$     (19)

Based on the data in Figures 7 (A) and (B), the estimated temperature is 424.43 nK and 450.82 nK, with a 6.22% difference. This small difference means that our algorithm does not destroy the original image information of the atomic cloud.

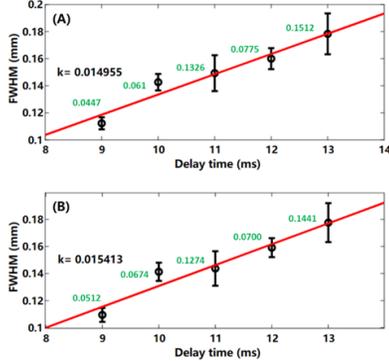

Fig. 7. The FWHM of atomic cloud at different delay times. (A) Conventional absorption imaging; (B) Image enhancement algorithm. Where the red solid line is the fitted straight line, and the green text is the width of the error bar.

Since noise influences statistical fluctuations in the particle population, the number of particles in the atomic cloud at different delay times can be used to test the algorithm's denoising impact. Comparing with that after algorithm improvement, the particle number recorded in the classic absorption imaging has higher error bar at the same delay time due to the unpredictability of the background noise, as illustrated in Figure 8, indicating that our method can greatly enhance particle number measurement accuracy.

As above, it verifies the image enhancement algorithm's high performance by measuring the size and quantity of atomic clouds. It can significantly minimize the impact of noise while keeping the atomic cloud's nature. Besides, it increases the contrast for atomic cloud photography and makes particle number monitoring easier.

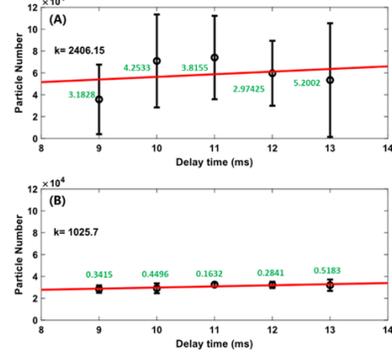

Fig. 8. The particles number of atomic cloud at different delay times. (A) Conventional absorption imaging; (B) Image enhancement algorithm. Where the red solid line is the fitted straight line, and the green text is the width of the error bar.

Our technique is resistant to complicated scenarios such as fringe-type noise and multi-component imaging. As illustrated in Figure 9 (A), fringe-type noise, such as global thick fringe and local thin fringe, might corrupt the image of the atomic cloud in various conditions. These fringes are difficult to remove by adjusting the optical path, but our technique can successfully handle their negative impacts (Figure 9 (B)). Furthermore, in multi-component imaging, our approach can still detect the local features of each component while providing outstanding de-noising (Figure 9 (D)), and the optimization procedure is totally automatic: no manually selected parameters are required.

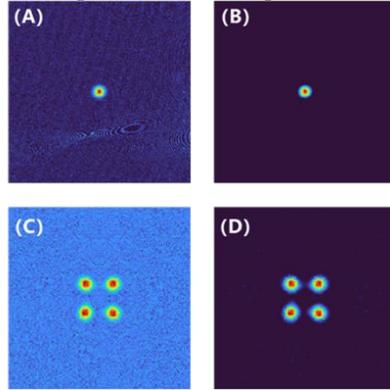

Fig. 9. Optimization results of fringe-type noise and multi-component imaging. (A) Absorption imaging image with fringe-type noise and its image enhancement results (B); (C) multicomponent absorption imaging and its image enhancement results (D).

## 5. Summary and Prospect.

In conclusion, we introduce an image enhancement algorithm for cold atom absorption imaging that progressively reduces noise and enhances image contrast through adaptive Gaussian filtering and nonlinear gray point operation. Experimental findings indicate that this approach can enhance particle population measurement accuracy by nearly tenfold without compromising the inherent characteristics of the atomic cloud. Moreover, even when confronted with fringe noise and multi-component imaging scenarios, the method consistently delivers exceptional results with high stability, and the optimization process is fully automated, eliminating the need for manual parameter selection. Our method offers greater operational flexibility and a broader range of applications compared to the previously published Optimized Fringe Removal Algorithm (OFRA) [40-41], as it calculates parameters based on a single image rather than requiring multiple photographs to construct a reference. This algorithm holds significant promise for advancing research in quantum control, precision measurement, and many-body quantum physics.

References.
1. K. S. Hardman, P. J. Everitt, G. D. Mcdonald, et al., "Simultaneous Precision Gravimetry and Magnetic Gradiometry with a Bose-Einstein Condensate: A High Precision, Quantum Sensor", Phys. Rev. Lett. 117, 138501.1-138501.5 (2016).
2. A. O. Jamison, B. Plotkin-Swing, S. Gupta, "Advances in precision contrast interferometry with Yb Bose-Einstein condensates", Phys. Rev. A 90, 063606 (2014).
3. A. D. Cronin, J. Schmiedmayer, D. E. Pritchard, "Optics and Interferometry with Atoms and Molecules", Rev. Mod. Phys. 81, 1051 (2009).
4. I. Bloch, J. Dalibard, W. Zwerger, "Many-Body Physics with Ultracold Gases", Rev. Mod. Phys. 80, 885-964 (2007).
5. M. Schreiber, S. S. Hodgman, P. Bordia, et al., "Observation of many-body localization of interacting fermions in a quasi-random optical lattice", Science 349, 842-845 (2015).
6. D. M. Farkas, K. M. Hudek, E. A. Salim, et al., "A compact, transportable, microchip-based system for high repetition rate production of Bose–Einstein condensates", Appl. Phys. Lett. 96, 091101 (2010).
7. G. W. Hoth, B. Pelle, S. Riedl, et al., "Point source atom interferometry with a cloud of finite size", Appl. Phys. Lett. 109, 071101 (2016).
8. D. A. Smith, S. Aigner, S. Hofferberth, et al., "Absorption imaging of ultracold atoms on atom chips", Opt. Express 19, 8471-8485 (2011).
9. C. F. Ockeloen, R. Schmied, M. F. Riedel, et al., "Quantum metrology with a scanning probe atom interferometer", Phys. Rev. Lett. 111, 143001 (2013).
10. M. Egorov, B. Opanchuk, P. Drummond, et al., "Measurement of s-wave scattering lengths in a two-component Bose-Einstein condensate", Phys. Rev. A 87, 053614 (2013).
11. M. Kitagawa, K. Enomoto, K. Kasa, et al., "Two-color photo association spectroscopy of ytterbium atoms and the precise determinations of s-wave scattering lengths", Phys. Rev. A 77, 012719 (2008).
12. M. G. Ries, A. N. Wenz, G. Zuern, et al., "Observation of Pair Condensation in the Quasi-2D BEC-BCS Crossover", Phys. Rev. Lett. 114, 230401 (2015).
13. P. Dyke, E. D. Kuhnle, S. Whitlock, et al., "Crossover From 2D to 3D in a Weakly Interacting Fermi Gas", Phys. Rev. Lett. 106, 105304 (2011).
14. A. D. Cronin, J. Schmiedmayer, D. E. Pritchard, "Optics and interferometry with atoms and molecules", Rev. Mod. Phys. 81, 1051-1129(2009).
15. J. Grond, U. Hohenester, I. Mazets, J. Schmiedmayer, "Atom impact of atom-atom interferometry with trapped Bose-Einstein condensates: interactions", New J. Phys. 12, 065036(2010).
16. K. P. Marzlin, W. Zhang, E. M. Wright, "Vortex Coupler for Atomic Bose-Einstein Condensates", Phys. Rev. Lett. 79, 4728-4731 (1997).
17. R. Dum, J. I. Cirac, M. Lewenstein, et al., "Creation of Dark Solitons and Vortices in Bose-Einstein Condensates", Phys. Rev. Lett. 80, 2972-2975 (1998).
18. E. L. Bolda, D. F. Walls, "Creation of Vortices in a Bose-Einstein Condensate by a Raman Technique", Phys. Lett. A 246, 32-36 (1998).
19. M. R. Matthews, B. P. Anderson, P. C. Haljan, et al., "Vortices in a Bose-Einstein Condensate", Phys. Rev. Lett. 83, 2498-2501 (1999).
20. K. Wei, Y. Fu, J. Yang, et al., "A Physics-based Noise Formation Model for Extreme Low-light Raw Denoising", IEEE, 2755-2764(2020).
21. G. E. Healey, R. Kondepudy, "Radiometric CCD camera calibration and noise estimation", IEEE Trans. Pattern Anal. Mach. Intell. 16, 267–276 (1994).
22. R. D. Gow, D. Renshaw, K. Findlater, et al., "A comprehensive tool for modeling CMOS image-sensor-noise performance", IEEE Trans. Electron Devices 54, 1321-1329 (2007).
23. H. Wach, E. R. Dowski Jr., "Noise modeling for design and simulation of computational imaging systems", Proc. SPIE 5438, 159–170 (2004).
24. R. L. Baer, "A model for dark current characterization and simulation", in Proceedings of Sensors, Cameras, and Systems for Scientific/Industrial Applications VII. SPIE 6068, 37-48 (2006).
25. M. V. Konnik, J. S. Welsh, "High-level numerical simulations of noise in CCD and CMOS photosensors: review and tutorial", https://arxiv.org/abs/1412.4031(December 11, 2014).
26. C. Leyris, A. Hoffmann, M. Valenza, et al., "Trap competition inducing RTS noise in saturation range in N-MOSFETs", in Noise in Devices and Circuits III, Proc. SPIE 5844, 41-51 (2005).
27. W. K, Y. Fu, Y. Zheng, et al., "Physics-based noise modeling for extreme low-light photography", IEEE Trans. Pattern Anal. Mach. Intell. 44, 8520-8537 (2021).
28. D. Coe, N. A. Grogin, "Readout Dark: Dark Current Accumulation During ACS/WFC Readout", ACS ISR 2 (2014).
29. X. H. Han, Y. W. Chen, Z. Nakao, "An ICA-based method for Poisson noise reduction", in Knowledge-Based Intelligent Information and Engineering Systems: 7th International Conference, KES 2003, Oxford, UK, September 2003, Proceedings, Part I, Springer Berlin Heidelberg, 7, 1449-1454 (2003).
30. B. Desai, U. Kushwaha, S. Jha, et al., "Image filtering-techniques algorithms and applications", Appl. GIS 7, 101 (2020).
31. G. Deng, L. W. Cahill, "An adaptive Gaussian filter for noise reduction and edge detection", 1993 IEEE Conference Record Nuclear Science Symposium and Medical Imaging Conference, IEEE, 1615-1619 (1993).
32. Y. Ohtake, A. G. Belyaev, H. P. Seidel, "Mesh Smoothing by

Adaptive and Anisotropic Gaussian Filter Applied to Mesh Normals", in VMV 2002, 203-210.
33. M. Nitzberg, T. Shiota, "Nonlinear image filtering with edge and corner enhancement", IEEE Trans. Pattern Anal. Mach. Intell. 14, 826–833 (1992).
34. A. Barron, J. Rissanen, B. Yu, "The minimum description length principle in coding and modeling", IEEE Trans. Inf. Theory 44, 2743-2760 (1998).
35. J. Rissanen, Stochastic Complexity in Statistical Inquiry (World Scientific, 1998).
36. T. Tasdizen, R. Whitaker, P. Burchard, et al., "Geometric surface processing via normal maps", ACM Trans. Graph. (TOG), 22(4), 1012-1033 (2003).
37. Y. Ohtake, A. G. Belyaev, H. P. Seidel, "Mesh Smoothing by Adaptive and Anisotropic Gaussian Filter Applied to Mesh Normals", in VMV, 2, 203-210, (2002).
38. G. Gómez, "Local Smoothness in terms of Variance: the Adaptive Gaussian Filter", in Proceedings of BMVC, 2000, 1-10, (2000).
39. F. Galland, A. Jaegler, M. Allain, et al., "Smooth contour coding with minimal description length active grid segmentation techniques", Pattern Recognit. Lett. 32, 721-730 (2011).
40. Z. Ma, C. Han, X. Jiang, et al., "Production of $^{87}$Rb Bose-Einstein condensate in an asymmetric crossed optical dipole trap", Chin. Phys. Lett. 38(10), 103701 (2021).
41. L. Niu, X. Guo, Y. Zhan, et al., "Optimized fringe removal algorithm for absorption images", Appl. Phys. Lett. 113, 141101 (2018).
42. J. G. D. Hester, M. M. Tentzeris, "Inkjet-printed flexible mm-wave Van-Atta reflectarrays: A solution for ultralong-range dense multitag and multisensing chipless RFID implementations for IoT smart skins", IEEE Trans. Microwave Theory Tech. 64, 4763-4773 (2016).

**Funding Sources.** This work is supported by the National Key Research and Development Program of China (2022YFA1404104), and the National Natural Science Foundation of China (12025509, 12104521).